\documentclass[%
 reprint,
superscriptaddress,
 amsmath,amssymb,
 aps,
]{revtex4-2}

\usepackage{graphicx}
\usepackage{dcolumn}
\usepackage{bm}
\usepackage{amsmath}
\usepackage{xcolor}
\usepackage[normalem]{ulem}

\begin{document}

\preprint{APS/123-QED}

\title{Capturing the slow relaxation time of superparamagnetic colloids in time-varying fields}

\author{Lucas H. P. Cunha}
\email{lh36@rice.edu}
\affiliation{Department of Chemical and Biomolecular Engineering, Rice University, Houston, Texas 77005, USA}%
\affiliation{Center for Theoretical Biological Physics, Rice University, Houston, TX 77005}%
\author{Aldo Spatafora-Salazar}
\email{astefanoss@rice.edu}
\affiliation{Department of Chemical and Biomolecular Engineering, Rice University, Houston, Texas 77005, USA}%
\author{Dana M. Lobmeyer}
\email{dml7@rice.edu}
\affiliation{Department of Chemical and Biomolecular Engineering, Rice University, Houston, Texas 77005, USA}%
\author{Kedar Joshi}
\email{kj18@rice.edu}
\affiliation{Department of Chemical and Biomolecular Engineering, Rice University, Houston, Texas 77005, USA}%
\author{Fred C. MacKintosh}
\email{fcmack@rice.edu}
\affiliation{Department of Chemical and Biomolecular Engineering, Rice University, Houston, Texas 77005, USA}%
\affiliation{Center for Theoretical Biological Physics, Rice University, Houston, TX 77005}%
\affiliation{Department of Physics and Astronomy, Rice University, Houston, TX 77005}%
\affiliation{Department of Chemistry, Rice University, Houston, TX 77005}%
\author{Sibani Lisa Biswal}
\email{biswal@rice.edu}
\affiliation{Department of Chemical and Biomolecular Engineering, Rice University, Houston, Texas 77005, USA}%


\date{\today}

\begin{abstract}
Superparamagnetic colloids present interesting assembly dynamics and propulsion in time-varying magnetic fields due to their magnetic relaxation. However, little is known about the mechanisms governing this magnetic relaxation, which is commonly attributed to the interactions and polydispersity of the ferromagnetic nanoparticles distributed within the colloid. We measure this relaxation from the effective potential between colloids subjected to rotating magnetic fields. Remarkably, our results indicate the presence of magnetic relaxation times much longer than what has been reported, which furthers our understanding of the magnetization of colloids in complex magnetic fields.

\end{abstract}

\maketitle

\begin{figure}
    \centering
    \includegraphics[width=0.48\textwidth]{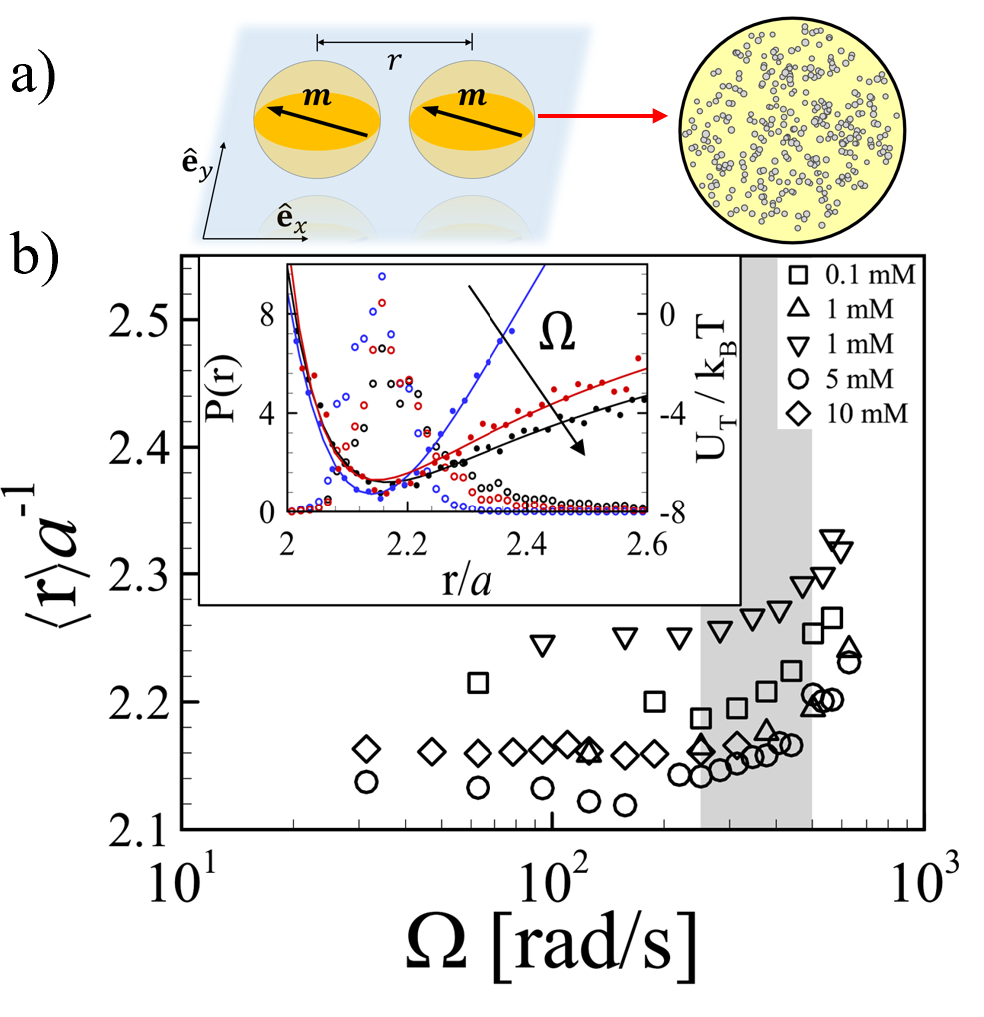}
    \caption{a) On the left, a pair of superparamagnetic colloids with magnetization $\textbf{m}$ separated by a center-to-center distance $r$ just above the glass surface. On the right, a not to scale representation of the ferromagnetic nano-particles (in gray) dispersed through the volume of the polymer matrix (in yellow) of the micron-sized superparamagnetic colloid. b) Time average of the beads' center-to-center distance, $\langle r \rangle$, as a function of $\Omega$ for different salt concentrations. The gray area in the plot represents the region where the product between the field frequency in rad/s, $\Omega$, and the beads' effective magnetic relaxation time in s, $\tau_m$, should be close to unity. In the inset, the empty symbols represent the probability distribution of $r$, $P(r)$, for a dimer subject to the field frequencies 50 Hz (314 rad/s, blue), 80 Hz (502 rad/s, red), and 100 Hz (628 rad/s, black), at the salt concentration of 5 mM. The solid symbols represent the respective dimer potential, $U_T \sim \log[P(r)]$, with the fitted curves obtained via Eq.~\ref{eq:dimer_mdm} (solid lines).}
    \label{fig:magentization}
\end{figure}

The active manipulation of superparamagnetic colloidal beads represents the basis for advanced microscale technologies, such as design of micro-robots and pumps \cite{biswal2004, du2018, dreyfus2005, yang2017, zimmermann2022, spatafora2023}, development of tunable materials \cite{morillas2020, alharraq2020, spatafora2022}, experimental modeling of molecular structures \cite{yethiraj2003, joshi2022, spatafora2021, lobmeyer2022, cunha2022}, and measurements with micro-rheological probes \cite{de2005, van2015}. However, the full realization of this application potential requires a better understanding of the magnetization of such colloids in rotating magnetic fields. Our approach here can lead to more accurate modeling of colloidal assembly dynamics in complex magnetic fields. Superparamagnetic colloids are structures composed of single-domain ferromagnetic nano-particles embedded in a solid nonmagnetic (diamagnetic) micron-sized polymer matrix, as illustrated in Fig.~\ref{fig:magentization}a \cite{fonnum2005}. When exposed to a magnetic field, the nano-particles tend to align their dipoles in the direction of the external field, whose collective response results in an effective magnetic dipole of the superparamagnetic colloid. For a given ferromagnetic domain, the relaxation time has been reported as $\mathcal{O}(10^{-9}\,s)$ \cite{tierno2007}. However, this relaxation time scale does not reflect the collective magnetization response of these nanoparticles embedded in the colloid \cite{fonnum2005,usov2020,janssen2009}. Recent experiments show that under rotating magnetic fields, superparamagnetic colloids are subjected to magnetic torques and experience an effective demagnetization as the field frequency increases \cite{janssen2009, du2013, coughlan2016, martinez2015, martinez2015b, massana2019}. Such phenomena are classical signatures of the presence of relaxation mechanisms with characteristic times comparable to the applied field.\cite{abdi2018, Sherman2019, Mignolet2022} Further complications arise when considering the pulsed fields associated with magnetic relaxivity. Various reports have shown that the relaxation time in pulsed magnetic fields can vary over six orders of magnitude for similar colloids, depending on the field intensity and duration \cite{liu2012, huang2017}.
 
Here, we utilize the fluctuations of the interparticle spacing between Brownian superparamagnetic colloids in rotating fields to determine their effective interaction potential.  An analytical expression for the potential is used to identify a relaxation time, which is considerably longer than previously reported \cite{martinez2015, coughlan2016, janssen2009}. For this, we introduce an experimental technique based on the determination of the effective potential between the beads in a dimer subject to a rotating magnetic field, but without the need for torque quantification. The method relies on a high enough field frequency where the particle interactions are time-averaged to be isotropic \cite{du2013, coughlan2017}. Specifically, the magnetization of the colloids is modeled using the classical Kelvin-Voigt model. Previous studies have indirectly captured relaxation effects of colloids in rotating magnetic fields by measuring the magnetic torque on individual beads \cite{coughlan2016,janssen2009,van2013}. Although the magnetic torques are a consequence of relaxation mechanisms, this approach does not provide direct measurements of the beads' magnetization, but rather the cross-product between the magnetization and the external field. A better understanding of the magnetization is needed to model colloidal assembly dynamics in complex fields. 

Polystyrene superparamagnetic colloidal beads with a nominal radius of $a=1.4\,\mu$m and volumetric magnetic susceptibility of $\chi=0.96$ (Dynabeads\textsuperscript{TM} M-270) are used for the experiments. The size of the colloids is necessary to guarantee observable thermal fluctuations with optical microscopy. The colloids are suspended in NaCl solutions of varying concentrations (0.1, 1, 5, and 10 mM) and placed inside a chamber made of two glass coverslips spaced approximately 100 $\mu$m apart. The density of the particles causes them to settle under gravity and the negative surface charge of the beads (-30mV) leads to repulsive electrostatic interactions with the glass slide.  The result is colloids that settle in a single plane, just above the glass surface \cite{li2011}. Hence, the dynamics of the system is confined to two dimensions \cite{behrens2001, tierno2007}. The colloid concentration is fixed at 0.01 mg/mL. A rotating magnetic field is induced using two pairs of air-core solenoids placed perpendicular to each other to generate a uniform magnetic field parallel to the glass slides. The field intensity was set to 12 Gauss, or 956 A/m, for all cases.  Colloidal dimers were typical for the conditions studied. 

Analyzing the time average of the colloids' center-to-center distance, $\langle r \rangle$, as a function of field frequency, $\Omega$, for different dimers (Fig.~\ref{fig:magentization}b), we observe the tendency for the colloids to separate from one another as $\Omega$ increases above a threshold value of about 250 rad/s. These results suggest weakening of the magnetic interactions between dimers at higher frequencies, indicative of an effective demagnetization of the colloids. Note that this threshold value $\Omega$ does not depend on the salt concentration. Hence, one might attribute the apparent demagnetization effects to internal magnetic relaxation mechanisms, which prevent the colloids from reaching their corresponding saturation magnetization. As will be further elaborated, the gray area in Fig.~\ref{fig:magentization}b represents the region in which $\Omega \tau_m \approx 1$, where $\tau_m$ is the colloids' magnetic relaxation time in s and $\Omega$ is given in rad/s. Brownian motion irreversibly separates the colloids when the field frequency is increased above the values shown in Fig.~\ref{fig:magentization}b.  The inductance of the solenoid system was tested to confirm that the demagnetizing effects did not arise from weakening of the external applied field.

The inset in Fig.~\ref{fig:magentization}b presents the probability distribution of $r$, $P(r)$, together with the dimer potential, $U_T(r)$, obtained from the Boltzmann distribution \cite{coughlan2017, du2013}. As $\Omega$ increases, the upper tail of the $P(r)$ distribution broadens. This elongation indicates that the beads spend a longer time away from each other, explaining the larger values of $\langle r\rangle$. The $U_T(r)$ shows a short-range repulsion due to electrostatics and a long-range attraction due to magnetic forces. The electrostatic-dominated region is not significantly affected by changes in $\Omega$, while the magnetic potential flattens as $\Omega$ increases. This result confirms that the increase in $\langle r\rangle$ with increasing $\Omega$ is due to weakening of the magnetic attraction (demagnetizing effect).

We model this decrease in internal magnetization with increasing frequency using the Kelvin-Voigt model with a single relaxation time $\tau_m$. For simplicity, we neglect any permanent magnetization or magnetic anisotropy that may arise from the heterogeneous distribution of the ferromagnetic nano-particles within the colloids or shape deviations from spherical \cite{coughlan2016,janssen2009}. Hence, the magnetization of the colloid can be described as:
\begin{equation}
\label{eq:mag_evolution}
\frac{\partial \textbf{m}}{\partial t} = -\frac{1}{\tau_m} [\textbf{m} - \chi_p (\textbf{H}_0 + \textbf{H}_I)] + \boldsymbol{\omega}_{p} \times \textbf{m},
\end{equation}
\noindent
where $\textbf{m}$ is the magnetic moment and $\chi_p = \frac{4}{3}\pi a^3\chi$ is the magnetic susceptibility of the colloid, $t$ is time, $\textbf{H}_0$ is the external time-varying magnetic field, $\textbf{H}_I = \sum_{i\neq j}{ \frac{\textbf{m}_j}{4\pi}\cdot \left( \frac{3\textbf{r} \textbf{r}}{r^5} - \frac{\textbf{I}}{r^3} \right) }$, which accounts for magnetic interactions between the $i^{th}$ and $j^{th}$ colloids with center-to-center distance $\textbf{r}$, $r=|\textbf{r}|$, $\textbf{I}$ is the identity tensor, and $\boldsymbol{\omega}_{p}$ is the angular velocity of each individual colloid. From Stokes' law, an individual non-Brownian particle rotates at $\boldsymbol{\omega}_p = \textbf{T}/8\pi\eta a^3$, where $\textbf{T} = \mu_0\textbf{m}\times\textbf{H}_0$ is the magnetic torque, and $\eta \approx 0.001$ Pa$\cdot$s is the solvent's viscosity. In this sense, we have that $\omega_p\sim \chi \mu_0 H_0^2/6 \eta$. For the experimental parameters used in this work, we estimate $\omega_p^{-1} = \tau_r\approx 5$ ms, which is the Brownian relaxation time of the colloid. Note that for $\tau_m \rightarrow 0$, Eq.~\ref{eq:mag_evolution} recovers the mutual-dipole-model, i.e., the colloid has no magnetic memory \cite{spatafora2021}

\begin{figure}
\centering
  \includegraphics[width=.5\textwidth]{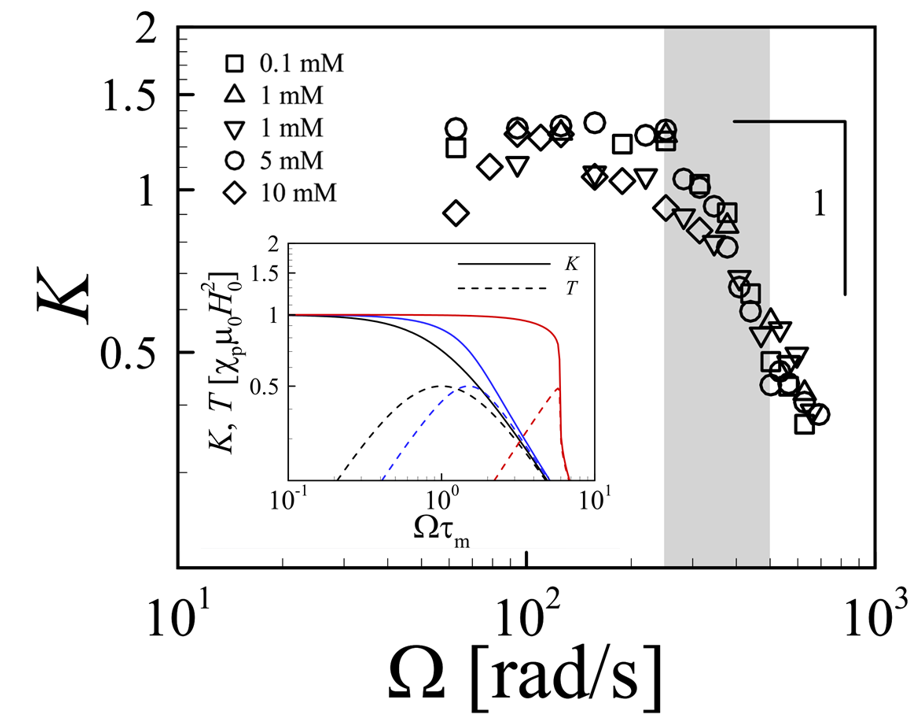}
  \caption{Experimental measurements of the demagnetizing factor $K$ as a function of the field frequency, $\Omega$, for different salt concentrations. Here, $K$ is obtained by fitting the potential curves using Eq.~\ref{eq:dimer_mdm}. The inset contains the theoretical values of $K$ and $T$ as a function of $\Omega \tau_m$ for an individual bead obtained by numerically solving Eq.~\ref{eq:mag_evolution} for $\tau_m/\tau_r = 0$ (black), $\tau_m/\tau_r = 1$ (blue), and $\tau_m/\tau_r = 10$ (red).}
\label{fig:K}
\end{figure}

The magnetic dipole of a superparamagnetic colloid in a rotating magnetic field is directed by a balance between the Néel relaxation, due to internal rotation within the colloid, and Brownian relaxation, due to the physical rotation of the colloid. The ratio $\tau_m/\tau_r$ defines how the magnetic dipole follows an external magnetic field.  Consider an isolated colloid subject to a rotating magnetic field, $\textbf{H}_0(t) = H_0[\cos(\Omega t)\hat{\textbf{e}}_x + \sin(\Omega t)\hat{\textbf{e}}_y]$, and assume $\tau_m/\tau_{r}\rightarrow 0$. In this scenario, the solution of  Eq.~\ref{eq:mag_evolution} shows that at long times ($t \gg \tau_m$), the colloid reaches a steady magnetization $m(\Omega) = \chi_p H_0 K$, where $K = [(\Omega^2 \tau_m^2 + 1)\cos(\phi)]^{-1}$, and $\phi = \tan^{-1}(\Omega\tau_m)$ is the phase lag angle between $\mathbf{m}$ and $\mathbf{H_0}$. We may interpret $K$ as the colloid's magnetization normalized by its saturated value for the given field strength. A quick analysis shows us that increasing $\Omega$ leads to weaker $K$ and larger $\phi$, two classical signatures of magnetic relaxation effects. Moreover, the isolated colloid becomes subject to $\mathbf{T} = \chi_p \mu_0 H_0^2 \frac{\Omega \tau_m}{\Omega^2 \tau_m^2+1} \, \hat{\textbf{e}}_z$.

The inset in Fig.~\ref{fig:K} presents the theoretical curves for $K$ and $T$ as a function of $\Omega$, for different ratios of $\tau_m/\tau_r$. For $\tau_m/\tau_r = 0$, the peak $T = \chi_p \mu_0 H_0 /2$ lies at $\Omega\tau_m = 1$. This peak shifts to higher frequencies as $\tau_m/\tau_r$ increases. At low frequencies, $K$ remains saturated at the unit value, while $T$ increases linearly, following the trend for $\phi$. In contrast, for $\Omega\tau_m \gg 1$, the relation inverts and $T \sim 1/\Omega\tau_m$. Although the single colloid rotates in response to magnetic torques, for $\tau_m/\tau_r \rightarrow 0$ this rotation plays no significant role on $\mathbf{m}$. However, as $\tau_m/\tau_r$ increases, the rotation of the colloid causes smaller values of $\phi$ and larger $m$. As a result, $T$ decreases with $\tau_m/\tau_r$ for lower frequencies but increases with $\tau_m/\tau_r$ for higher frequencies. Note that at $\Omega\tau_m = 1$, the colloid's magnetization already presents an appreciable decay for $\tau_m/\tau_r=0$, $K\approx 0.7$. The grey region in Fig.~\ref{fig:magentization}b represents where $\Omega\tau_m = 1$ and correspondingly, $2\leq\tau_m\leq 4$ ms. Additionally, the ratio $\tau_m / \tau_r$ can be estimated to be between $0.37$ and $0.74$ for our experiments. From the small deviations between the theoretical curves of $K(\Omega)$ for $\tau_m / \tau_r = 0$ and 1 in the inset of Fig.~\ref{fig:K}, we proceed our analyses assuming no significant effects arise from Brownian relaxation mechanisms. 

Assuming that the field frequency is high enough such that the translation of the colloids is negligible in a single cycle, the magnetic potential between two colloids is effectively isotropic and can be described using the mutual-dipole-model \cite{du2013} 
\begin{equation}
    \label{eq:dimer_mdm}
    \bar{U}_{m} = \frac{2\pi a^3 \mu_0 H_0^2 \tilde{r}^3}{9} \left[ \frac{\bar{\chi}^2}{\left(\tilde{r}^3+\bar{\chi}/3\right)^2} - \frac{2 \bar{\chi}^2}{\left(\tilde{r}^3-2\bar{\chi}/3\right)^2} \right],
\end{equation}
\noindent
for $\tilde{r} = r/a$. In an attempt to take into account the magnetic relaxation effects, we set $\bar{\chi} = \chi K(\Omega\tau_m)$ as an effective magnetization for a single colloid. Such a model is simpler to consider when compared to the solution of Eq.~\ref{eq:mag_evolution} for a pair of interacting colloids,
\begin{eqnarray}
    \label{eq:dimer_complete}
    \bar{U}_{m} = \frac{2\pi a^3 \mu_0 H_0^2}{9\tilde{r}^3} & \left[ \frac{\chi^2 \Omega^2\tau_m^2}{\left(\tau_m^2\Omega^2+\gamma_y^2\right)^2\sin^2(\phi_y)} \right. \nonumber \\ 
     & \left. -\frac{2\chi^2 \Omega^2\tau_m^2}{\left(\tau_m^2\Omega^2+\gamma_x^2\right)^2\sin^2(\phi_x)} \right]
\end{eqnarray}
\noindent
where $\phi_{x,y} = \arctan{(\tau_m\Omega/\gamma_{x,y})}$, $\gamma_{x} = 1 - 2\chi/3\tilde{r}^3$, and $\gamma_{y} = 1 + \chi/3\tilde{r}^3$. In both Eqs.~\ref{eq:dimer_mdm} and~\ref{eq:dimer_complete}, we assumed $\tau_{m}/\tau_r \rightarrow 0$. Direct comparisons between both solutions give that Eq.~\ref{eq:dimer_mdm} underestimates the magnitude of $\bar{U}_{m}$ when compared to Eq.~\ref{eq:dimer_complete}, presenting a maximum error of about $10\%$ when $\Omega\tau_m \approx 2$ for colloids in contact. 

To account for the electrostatic repulsive interactions between the colloids, we consider the potential $U_e = C\exp{[-\kappa a(\tilde{r}-2)]}$, where $\kappa^{-1}$ is the Debye length and $C$ is a constant for the purposes of the present work \cite{berg2010}. Thus, the total potential between the dimer becomes $U_T = \bar{U}_{m} + U_e$. Finally, using Eqs.~\ref{eq:dimer_mdm} and~\ref{eq:dimer_complete} to fit the data for $U_T(\tilde{r})$ obtained by the Boltzmann distribution we measure $K(\Omega)$ and $\tau_m$, respectively. Examples of the curve fit for $U_T(\tilde{r})$ using Eq.~\ref{eq:dimer_mdm} are shown in the inset of Fig.~\ref{fig:magentization}b.   

\begin{figure}
\centering
  \includegraphics[width=.5\textwidth]{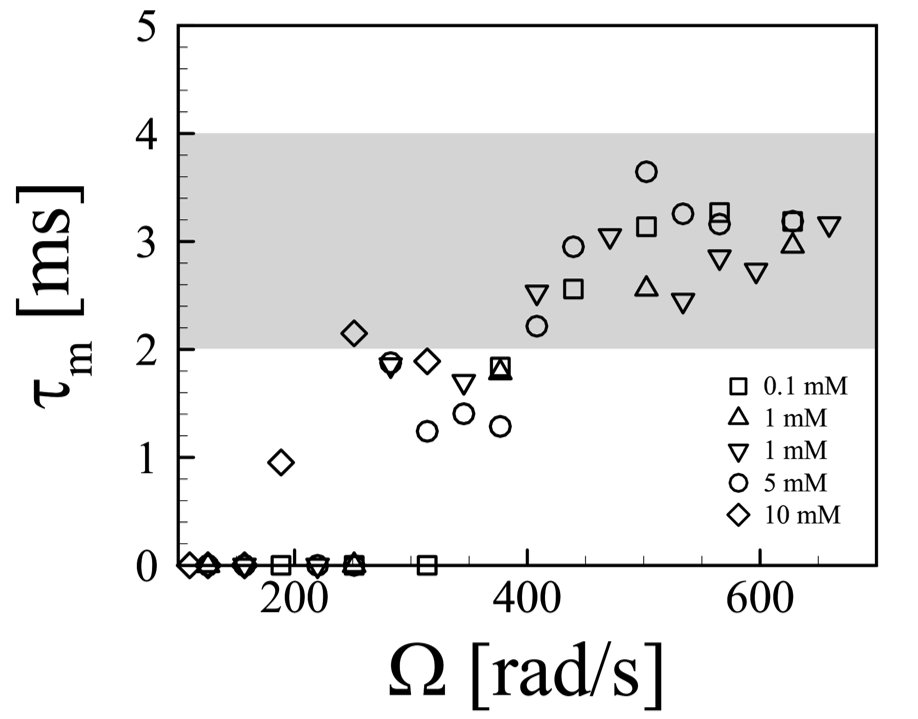}
  \caption{Magnetic relaxation time, $\tau_m$, as a function of $\Omega$ obtained from the curve fit considering Eq.~\ref{eq:dimer_complete}. The gray regions refer to the region in which we speculate $\Omega\tau_{m} = 1$ to be, according to the analysis from Figs.~\ref{fig:magentization} and~\ref{fig:K}.}
\label{fig:tau_mag}
\end{figure}

It is possible to obtain $\bar{\chi}$ from fitting the experimental data to $\bar{U}_T$ using  Eq.~\ref{eq:dimer_mdm}, with known experimental parameters $2\pi a^3 \mu_0 H_0^2 / 9k_B T = 532.5$ and $\chi=0.96$. Figure~\ref{fig:K} presents the measured curves of $K$ as a function of $\Omega$ for dimers at different salt concentrations. All cases behave similarly. For $\Omega\lessapprox\,250$ rad/s, $K$ saturates to a value around 1.2. This saturation implies that the beads are fully magnetized and no magnetic relaxation effect is appreciable in this regime. The values of $K$ greater than unity may be attributed to the higher order interactions between the colloids not taken into account in the mutual dipolar interaction model used for Eq.~\ref{eq:dimer_mdm} \cite{du2014,spatafora2021}. Increasing $\Omega$ leads to the gray region, where $\Omega\tau_m \approx 1$ from the previous analysis. Here, the relaxation effects become evident from the decreasing $K$ values. For $\Omega \gtrapprox 400$ rad/s, $K$ approximately scales with $\Omega^{-1}$, agreeing with the theoretical predictions shown in the inset of Fig.~\ref{fig:K}. 

Proceeding with Eq.~\ref{eq:dimer_complete} for the interaction potential curve fits, direct measurements of $\tau_{m}$ are obtained. Additionally, from the experimental parameters, $2\pi a^3 \mu_0 H_0^2 \chi^2/ 9k_B T = 490$. Figure~\ref{fig:tau_mag} presents $\tau_{m}$ as a function of $\Omega$ for different salt concentrations. One observes that for lower frequencies, the measured $\tau_m$ is virtually zero.  Although this regime is characterized by isotropic interaction potential,the relaxation mechanisms are not captured by the method. However, at higher frequencies relaxation mechanisms induce observable effects, i.e., $\Omega \gtrapprox \tau_{m}^{-1}$. Notably, for $\Omega > 400$ rad/s, we measure effective relaxation times in agreement with the values speculated from the previous analyses (the gray region between 2 ms and 4 ms),  corroborating the consistency of the methodology.  

\begin{figure}
\centering
  \includegraphics[width=.5\textwidth]{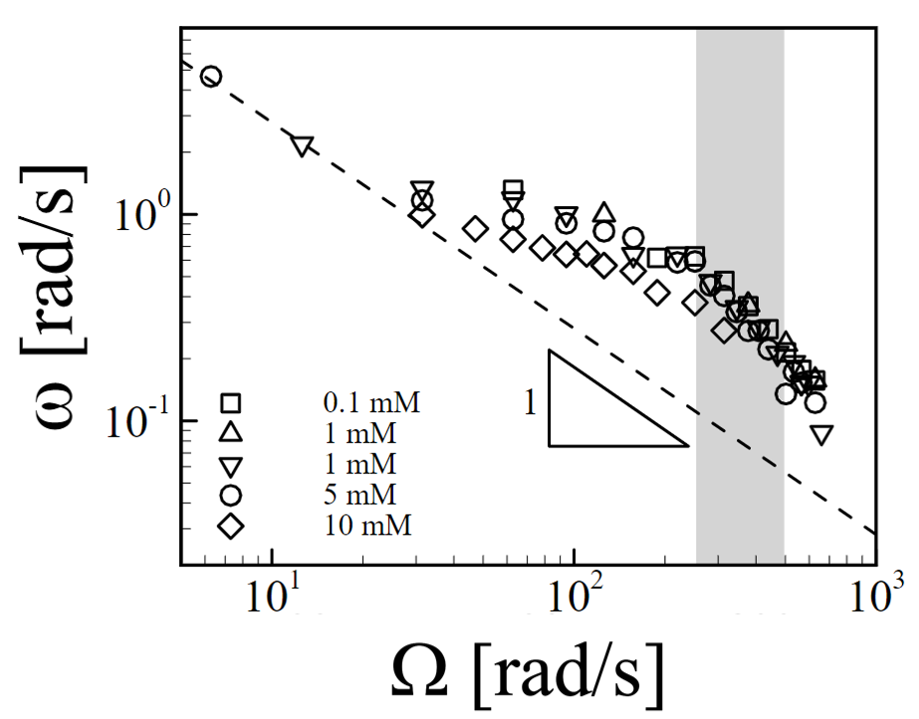}
  \caption{Dimer rotation frequency, $\omega$, as a function of the field frequency, $\Omega$, for different salt concentrations. The dashed line refers to the trend expected in the absence of magnetic relaxation effects ($\tau_{m}\rightarrow 0$), described by the asynchronous dynamics regime. The gray region refers to the region in which we speculate $\Omega\tau_{m} = 1$ to be, according to the analysis from Figs.~\ref{fig:magentization} and~\ref{fig:K}.}
\label{fig:rotation}
\end{figure}

Besides the apparent demagnetizing effect, typical relaxation mechanisms often manifest as a delay of the colloids' magnetization with respect to the applied magnetic field, causing magnetic torques to act on the individual particles. This phenomenon of torqued colloids has been used for the controlled transport of cargo in micro-systems, e.g., the carpets and micro-wheels from Massana et. al \cite{massana2019} and Zimmermann et al. \cite{zimmermann2022}, respectively. The torque-induced rotation of the magnetic colloids produces circulating flows around each of them, leading to their propulsion when hydrodynamically interacting with near walls, analogous to wheels on the road. For a pair of colloids, the hydrodynamic coupling between the particles causes the dimer to rotate around its center. Figure~\ref{fig:rotation} presents the dimers rotating frequency, $\omega$, as a function of $\Omega$. 

In the absence of torques on individual colloids, dimers rotate following either synchronous dynamics at low frequencies or asynchronous dynamics at high frequencies \cite{biswal2004, coughlan2016, tierno2007, spatafora2021}. When in the synchronous regime, the dimers rotate with the same frequency as the rotating field. While in the asynchronous regime, the dimers rotate with a frequency inversely proportional to the field one, a trend which is captured by the dashed line in Fig.~\ref{fig:rotation}. The bump observed in the $\omega(\Omega)$ curve with respect to the dashed line around $\Omega\tau_m\approx 1$ highlights the contribution of the magnetic torques to the dimers' rotation due to their hydrodynamic coupling. Such a result validates the theoretical prediction for the magnetic torque shown in the inset of Fig.~\ref{fig:magentization}, where $T$ is maximized at $\Omega\tau_m=1$. Notably, both Massana et. al \cite{massana2019} and Zimmermann et al. \cite{zimmermann2022} used field frequency of $250$ rad/s to drive their carpets and micro-wheels which are assembled with the same superparamagnetic colloidal beads used in this work, showing consistency between the present study and previous results with collective colloids.

In this work, we present a model that captures the magnetic relaxation times of superparamagnetic colloids in rotating magnetic fields. For this, we introduced a methodology to measure the effective internal relaxation time of the superparamagnetic beads without the need for torque measurements. The model is based on the effective isotropic potential between dimers subject to a rotating magnetic field. For a magnetic field strength of $H_0 = 956$ A/m, we find that the particles' magnetic relaxation time is between 2 to 4 ms, at least one order of magnitude greater than what was previously reported in the literature for different field strengths for similar particles in rotating magnetic fields\cite{martinez2015, massana2019}. Furthermore, we show that for $\Omega\tau_m \approx 1$, the magnetic torques on individual colloids hydrodynamically induce faster rotation of the dimers. Our results can be valuable for active microrheology and may aid in the design of micro-robots \cite{janssen2009,lipfert2015,behrens2001,van2015,van2013, zimmermann2022,massana2019}. Moreover, these magnetic relaxation effects can be used to induce more complex interactions between colloids. Deeper investigations are required to better understand the nature of the observed relaxation times. We suspect that it may be a result of magnetic frustrations in clusters of the ferromagnetic nano-particles commonly found inside the colloids \cite{fonnum2005}. 

\section*{Acknowledgments}
We thank Dr. Cristiano Nisoli for the insightful discussions on magnetic relaxation mechanisms. This work was supported in part by the National Science Foundation Divisions of Materials Research (Grant No. DMR-2224030), Center for Theoretical Biological Physics (Grant No. PHY-2019745), and Directorate for Technology, Innovation, and Partnerships (Grant No. PFI-2141112).

\bibliography{apssamp}

\end{document}